\documentclass{article}
\usepackage{amsmath,graphicx,mlspconf,amssymb}
\usepackage{microtype}
\usepackage{mlspconf}
\usepackage{url}
\usepackage{siunitx}
\usepackage{booktabs}
\usepackage{multirow}
\usepackage[skip=0.8\baselineskip]{caption}
\let\OLDthebibliography\thebibliography
\renewcommand\thebibliography[1]{
  \OLDthebibliography{#1}
  \setlength{\parskip}{0pt}
  \setlength{\itemsep}{0.4pt plus 0.1ex}
}
\copyrightnotice{978-1-7281-6338-3/21/\$31.00 {\copyright}2021 Crown}

\toappear{2021 IEEE International Workshop on Machine Learning for Signal Processing, Oct.\ 25--28, 2021, Gold Coast, Australia}

\title{Conditional Sound Generation Using Neural Discrete Time-Frequency Representation Learning}
\name{Xubo Liu$^1$, Turab Iqbal$^1$, Jinzheng Zhao$^1$, Qiushi Huang$^2$, Mark D. Plumbley$^1$, Wenwu Wang$^1$}
\address{
  $^1$Centre for Vision, Speech and Signal Processing (CVSSP), University of Surrey, UK,\\
  \{xubo.liu\thanks{This work was partly supported by the DASA project MANTIS Phase 2, a Research Scholarship from the China Scholarship Council No. 202006470010 and a PhD Studentship from the University of Surrey.}, t.iqbal, j.zhao, m.plumbley, w.wang\}@surrey.ac.uk\\
  $^2$Department of Computer Science, University of Surrey, UK, qiushi.huang@surrey.ac.uk
  }

\begin{document}

\maketitle

\begin{abstract}
Deep generative models have recently achieved impressive performance in speech and music synthesis. However, compared to the generation of those domain-specific sounds, generating general sounds (such as siren, gunshots) has received less attention, despite their wide applications. In previous work, the SampleRNN method was considered for sound generation in the time domain. However, SampleRNN is potentially limited in capturing long-range dependencies within sounds as it only back-propagates through a limited number of samples. In this work, we propose a method for generating sounds via neural discrete time-frequency representation learning, conditioned on sound classes. This offers an advantage in efficiently modelling long-range dependencies and retaining local fine-grained structures within sound clips. We evaluate our approach on the UrbanSound8K dataset, compared to SampleRNN, with the performance metrics measuring the quality and diversity of generated sounds. Experimental results show that our method offers comparable performance in quality and significantly better performance in diversity.
\end{abstract}

\begin{figure*}
  \centering
  \includegraphics[width=\linewidth]{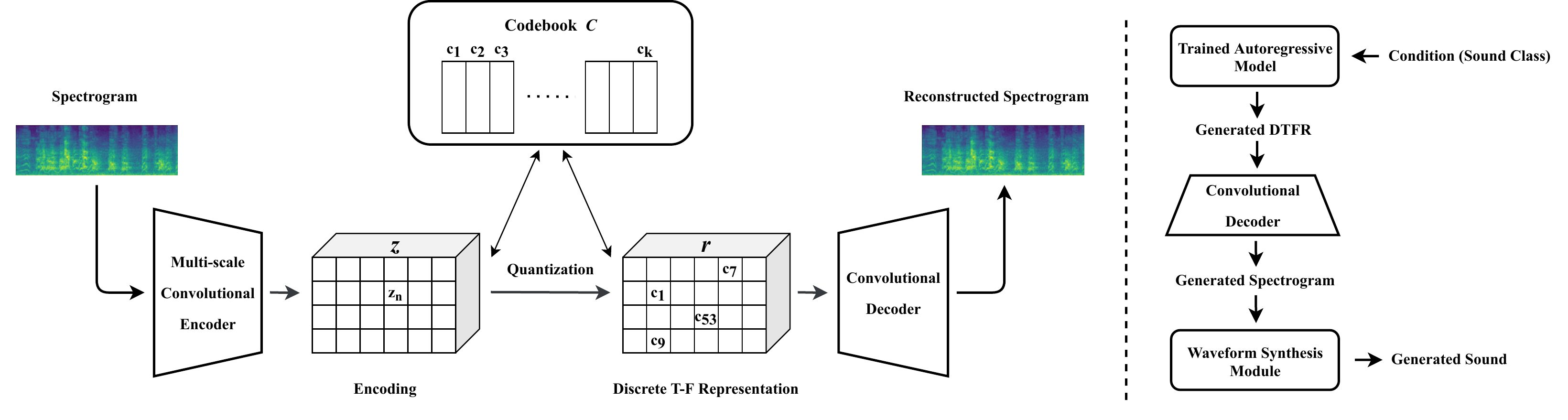}
  \caption{Left: The proposed VQ-VAE based approach to learn a discrete T-F representation (DTFR) of sound. Right: The pipeline for conditional sound generation in the inference stage. We train the VQ-VAE model and the autoregressive model separately.}
  \label{fig:vqvae}  
\end{figure*}

\section{Introduction}
\label{sec:intro}
General sounds carry a wide range of information about environments, from individual physical events to sound scenes as a whole \cite{virtanen2018computational}. General sound generation has many potential applications, such as the automatic production of sound effects for movies and video games \cite{lloyd2011sound}. In addition, due to the difficulties of collecting and annotating audio data, sound generation can be used as an efficient data augmentation \cite{salamon2017scaper} approach for acoustic scene classification \cite{ren2018deep} and sound event detection \cite{kong2020sound}. In the long term, sound search engines \cite{akkermans2011freesound} could incorporate a sound generation system and customize sound according to the personal tastes of users.

Recently, significant progress has been made in speech synthesis \cite{shen2018natural, wang2017tacotron} and music generation \cite{yang2017midinet, dieleman2018challenge} using deep generative models. Compared with domain-specific sounds such as speech and music, general sound is less structured and has greater diversity, typically accompanied by noise and reverberation. Therefore, it is challenging to model general sounds using deep generative models. Related work on general sound generation includes acoustic scene generation \cite{kong2019acoustic} and environmental sound synthesis \cite{okamoto2021onoma}. However, general sound generation remains a relatively unexplored area. 

SampleRNN \cite{mehri2016samplernn} is an autoregressive model for waveform generation, which has been adapted to sound generation by Kong et al. \cite{ kong2019acoustic}. SampleRNN generates sound in the time domain and only back-propagates through a fraction of a second \cite{vasquez2019melnet}. Thus, it is difficult to capture the long-range dependencies within sound clips using SampleRNN. However, some sound events typically have long-range dependencies, such as an ambulance siren spanning several seconds (tens of thousands of audio samples), and capturing these dependencies would be beneficial for the generation of such sounds. 

Modeling sound in the time-frequency (T-F) domain, e.g. using spectrogram, can help capture long-range dependencies \cite{vasquez2019melnet}, although an additional step is required to convert the T-F representation into a time domain waveform. Recently, GAN-based methods \cite{kumar2019melgan, kong2020hifi} have been proposed for waveform synthesis due to the computational efficiency offered by their parallel structure and good quality of synthesized audio. Synthesizing high-quality waveforms would normally require the spectrograms to be in high temporal resolution in order to retain the local and fine-grained characteristics that are important for sound fidelity. However, increasing the temporal resolution of the spectrogram (i.e., decreasing the short-time Fourier transform (STFT) hop size) would incur a higher computational cost.

In this paper, we propose an approach to generate sound conditioned on different sound classes in the T-F domain using a Vector Quantised Variational AutoEncoder (VQ-VAE) \cite{oord2017neural}. Our approach can model the long-range dependencies of sound while reducing the computational cost of modeling sound with high temporal resolution in the T-F domain. More specifically, a VQ-VAE model is trained to learn a discrete T-F representation (DTFR) of sound. Then, an improved autoregressive model \cite{chen2018pixelsnail} is trained using the DTFR as input and the sound classes as conditions to generate sounds. In addition, we propose a multi-scale convolutional scheme for the encoder of the VQ-VAE to capture acoustic information (i.e. features) of sound at different scales. We show that this leads to a compact DTFR while enables the encoding of local fine-grained structures of sound. To our knowledge, the VQ-VAE model has not yet been considered for the conditional generation of general sounds. We demonstrate empirically that our approach offers advantages in modeling the long-range dependencies of sound over the time-domain generation method \cite{ kong2019acoustic}.




We evaluate the diversity \cite{kong2019acoustic} and quality \cite{richardson2018gans} (as described in Section 3.5) of the generated sound samples on the UrbanSound8K dataset \cite{salamon2014dataset}. Experimental results show that our proposed method outperforms the SampleRNN baseline \cite{kong2019acoustic} in diversity and has comparable performance in quality. The code and generated samples are available on GitHub\footnote{\url{https://github.com/liuxubo717/sound_generation}}.

\section{Approach}
\label{sec:approach}


To generate sound conditionally, we first use a VQ-VAE \cite{oord2017neural} to learn a DTFR of sound, as described in Section 2.1. Then, the process of generating sound using the DTFR conditioned on sound class labels is summarized in Section 2.2.

\subsection{Discrete time-frequency representation learning}

To disentangle the spectrogram representation of sound into a compressed DTFR, we employ a VQ-VAE-based model consisting of an encoder, a decoder and a codebook. Our proposed approach assumes fixed input size, the encoder learns a non-linear mapping from the spectrogram $x \in \mathbb{R}^{H \times W \times 1}$ onto a latent representation $z \in \mathbb{R}^{H/2^m \times W/2^m \times D}$ ($H$, $W$, $D$ are height, width and depth, respectively), where $m$ is a compression factor. The latent representation $z$ consists of $N$ elements $z_n\in \mathbb{R}^{1 \times 1 \times D}$, where $N = H/2^m \times W/2^m$. Each element ${z_n}$ is quantized based on its distance to the codewords $c_k$ in the codebook $C=\{c_k\}_{k=1}^K$ with $K$ being the number of codewords in the codebook $C$. Formally:
\begin{equation}
  \operatorname{Quantize}(z_n) = c_k \ \text{where}\ 
  k = \mathop{\arg\min}_{i}\| z_n - c_i\|_2^{},
  \label{eq1}
\end{equation}
where $z_n$ is reshaped to a vector of the same dimension as $c_i$ for calculation. After the quantization of each element in $z$, the DTFR defined as $r=\{r_n\}_{n=1}^N$ is obtained, and is fed into the decoder to reconstruct the spectrogram. The reconstructed spectrogram $\hat{x}$ is given by:
\begin{equation}
  \hat{x} = \operatorname{Decoder}(r) = \operatorname{Decoder}(\operatorname{Quantize}(z)).
  \label{eq2}
\end{equation}

To learn the reconstruction process in Equation (\ref{eq2}), the gradient is passed from the decoder input to the encoder output. The loss function of the VQ-VAE is defined as follows:
\begin{equation}
\begin{aligned}
\mathrm{Loss} = \| x - \hat{x}\|_2^2 \ + \  \| \text{sg}[z] - r\|_2^2 + \ \beta\| \text{sg}[r] - z\|_2^2,
\label{eq3}
\end{aligned}
\end{equation}
where sg[·] denotes the stop-gradient operation \cite{oord2017neural}, which ensures the operand is not updated during backpropagation, and $\beta$ is a regularization parameter. The first term is a reconstruction loss, the second term is used to align the codebook with the encoder output, and the last term is a commitment loss \cite{oord2017neural}, which mitigates the uncertainty caused by noise in the mapping between the encoder output and the codewords.

\subsubsection{Multi-scale convolutional scheme in the encoder}
A conventional VQ-VAE uses a fully-convolutional encoder with a fixed kernel size, which can capture the local characteristics in the spectrograms but cannot make use of the dependencies between long-term temporal frames. To efficiently capture both local characteristics and long-range dependencies, we propose a multi-scale convolutional scheme in the encoder of the VQ-VAE. In this scheme, multi-scale CNNs with varied kernel sizes are deployed. This multi-scale convolutional approach has been shown to be effective in capturing the global and local information of audio signals in the T-F domain \cite{xian2021multi}.

More precisely, the encoder consists of several strided convolutional layers (SCLs) in parallel. Each SCL has several consecutive sub-layers with strided convolutional kernels of fixed sizes followed by residual blocks. These SCLs have different kernel sizes. SCLs with small kernels are used to capture the local characteristics between the adjacent temporal frames, and SCLs with large kernels are utilized to explore the dependencies between long-range temporal frames. Then, the output of each SCL is added together to obtain the output of the encoder, thus enabling the encoder to capture global and local information (i.e. acoustic features) at different scales. 

\subsubsection{Model architecture}
A fully-convolutional decoder is used to decode the DTFR to the reconstructed spectrogram. The structure of the decoder is similar to the encoder, except that the multi-scale convolutional scheme is omitted. The architecture of the proposed approach to learn the DTFR of sound is shown in Figure \ref{fig:vqvae} (left). Details of the model will be discussed in Section 3.3.

\subsection{Conditional sound generation}
After learning the DTFR of sound, the task of conditional sound generation can be treated as generating the DTFR of sound, conditioned on the class labels. Since the DTFR is a compressed and compact representation, we can significantly alleviate the computational cost of modeling sound while still retaining the long-range dependencies and local characteristics of the sound. The decoder of the trained VQ-VAE model in Section 2.1.2 is used to map the generated DTFR to the generated spectrogram. The generation of the DTFR of a sound is described as below.

Considering that the index $k$ of the codewords $c_k$ can characterise the $n$th component of any DTFR $r=\{r_n\}_{n=1}^N$ (as described in Section 2.1), we first formulate $r$ as a sequence of indexes $y=\{y_n\}_{n=1}^N$ as follows:
\begin{equation}
y_n = k \ \ \text{where} \ \ r_n = c_k.
  \label{eq4}
\end{equation}
Then, we use an autoregressive model to build the distribution $p(y)$ over the DTFR of sound by factorising the joint distribution as a product of conditionals:  
\begin{equation}
p(y) = p(y_1, ... , y_n) = \prod_{i=1}^{n}p(y_i|y_1, ... , y_{i-1}).
  \label{eq5}
\end{equation}
To generate sound conditioned on a class label, we apply the one-hot encoding vector $h$ of a sound class as the global condition. Formally:

\begin{equation}
p(y|h) = p(y_1, ... , y_n|h) = \prod_{i=1}^{n}p(y_i|y_1, ... , y_{i-1}, h).
  \label{eq6}
\end{equation}
We use PixelSNAIL \cite{chen2018pixelsnail} to build $p(y|h)$. PixelSNAIL is an improved autoregressive model that combines causal convolutions \cite{oord2016conditional} with self-attention \cite{vaswani2017attention}. After training the VQ-VAE, we compute the DTFR of sound using the encoder of the trained VQ-VAE. Then PixelSNAIL is trained on the DTFR conditioned on class labels. The generation of the new DTFR is enabled by sampling the variables conditioned on all previous variables one by one from the trained autoregressive model. A waveform synthesis module, namely HiFi-GAN \cite{kong2020hifi} (as described in Section 3.3.3), is deployed for converting the generated spectrogram into a waveform. The generation pipeline for our proposed approach in the inference stage is shown in Figure \ref{fig:vqvae} (right). 

\section{Experiments}
\label{sec:experiment}

\begin{table*}[t]
\caption{Results of classification accuracy}
\label{tab:quality}
\resizebox{\textwidth}{10.5mm}{
\begin{tabular}{@{}cccccccccccc@{}}
\toprule
                  & air\_conditioner & car\_horn & children\_playing & dog\_bark & drilling & engine\_idling & gun\_shot & jackhammer & siren & street\_music & \textbf{Average} \\ \midrule
Proposed Approach & 0.8516               & 0.5049        &  0.1738                 &  0.6875         &  0.9453        & 0.1875              & 0.7832         & 0.1240          & 0.6699     & 0.3613             & \textbf{0.5289}       \\ \midrule
SampleRNN & 0.6328               & 0.7119         & 0.7002                 & 0.3438         & 0.3984        & 0.2305              & 0.4980         & 0.5840          & 0.6191     & 0.5625             & \textbf{0.5281}       \\ \midrule
Test & 0.4400                &  0.9375         &  0.9200                 &  0.8500         & 0.6100        &  0.7640              & 0.9032         &  0.9146          & 0.9878     & 0.9700             & \textbf{0.8297}       \\ \midrule
Reconstructed Test         &  0.3500               & 0.9688         & 0.8100              &  0.8700        & 0.8300        & 0.5843           & 0.8710         & 0.9024          &  0.9878    & 0.9000           & \textbf{0.8074}    \\ \bottomrule
\end{tabular}}
\end{table*}


\begin{table*}[t]
\caption{Results of class-wise NDB and JSD}
\label{tab:diversity_class}
\resizebox{\textwidth}{14mm}{
\begin{tabular}{@{}ccccccccccccc@{}}
\toprule
\multicolumn{2}{c}{}                          & air\_conditioner & car\_horn & children\_playing & dog\_bark & drilling & engine\_idling & gun\_shot & jackhammer & siren & street\_music & \textbf{Average} \\ \midrule
\multirow{2}{*}{Proposed Approach} & NDB$_{\text{class}}$ &6	&4	&4	&2	&3	&1	&1	&3	&4	&3	&\textbf{3.1}       \\ \cmidrule(l){2-13} 
                                   & JSD$_{\text{class}}$ &0.0694	&0.0522	&0.0714	&0.0351	&0.0425	&0.0336	&0.0364	&0.0357	&0.0568	&0.0448	&\textbf{0.0478}   \\ \midrule
\multirow{2}{*}{SampleRNN}         & NDB$_{\text{class}}$ &15	&10	&11	&9	&11	&16	&8	&10	&12	&13	&\textbf{11.5}   \\ \cmidrule(l){2-13} 
                                   & JSD$_{\text{class}}$ &0.2897	&0.1748	&0.4859	&0.3130	&0.1632	&0.3017	&0.2856	&0.1363	&0.3251	&0.2955	&\textbf{0.2771}   \\ \midrule 
\multirow{2}{*}{Test} & NDB$_{\text{class}}$ &1	&1	&1	&0	&0	&2	&0	&1	&2	&0	&\textbf{0.8}      \\ \cmidrule(l){2-13} 
                                   & JSD$_{\text{class}}$ &0.2932	&0.1881	&0.1045	&0.0427	&0.0700	&0.3476	&0.2202	&0.3677	&0.2983	&0.0964	&\textbf{0.2029}   \\ \bottomrule                                   
\end{tabular}}
\vspace{-0.5em}
\end{table*}

\subsection{Dataset}
We evaluate our proposed approach for conditional sound generation on the UrbanSound8K dataset \cite{salamon2014dataset}. UrbanSound8K consists of \num{8732} labeled sound clips of urban sound from \num{10} classes. The duration of each sound clip is less than \num{4} seconds. UrbanSound8K has a large diversity of sound classes, such as siren and street music. In addition, each sound clip is divided into foreground sound or background sound. These attributes make it appropriate for using UrbanSound8K to evaluate the ability of the generative model to capture the salient features of different sound classes. UrbanSound8K is divided into \num{10} folds and we use the predefined folds to obtain \num{7916} sound clips for training and \num{816} sound clips for testing. Because our proposed approach assumes that the length of input audio is fixed, we pad all sound clips to \num{4} seconds. All sound clips are converted to \SI{16}{bit} and down-sampled to \SI{22050}{\kHz}.

\subsection{Spectrogram computation}
To generate high quality sound, we compute the spectrogram with the hyperparameter values as used in HiFi-GAN \cite{kong2020hifi}, which can achieve high-fidelity waveform synthesis, as described in Section 3.3.3. More precisely, we use an \num{80}-dimensional log mel-spectrogram calculated using the short-time Fourier transform (STFT) with a frame size of \num{1024}, a hop size of \num{256}, and a Hann window. Dynamic range compression is applied to the mel-spectrogram by first clipping it to a minimum value of \num{1e-5} and then applying a logarithmic transformation. A sound clip of 4 seconds results in a mel-spectrogram with shape $80 \times 344$.

\subsection{Details of model implementation}
\subsubsection{VQ-VAE}
For the encoder of the VQ-VAE, we use four SCLs consisting of two sub-layers with stride \num{2}, followed by two $3 \times 3$ residual blocks (ReLU, $3 \times 3$ conv, ReLU, $1 \times 1$ conv). The kernel sizes of these four SCLs are $2 \times 2$, $4 \times 4$, $6 \times 6$ and $8 \times 8$ respectively. Thus, we can down-sample the input log mel-spectrogram from $80 \times 344$ to $20 \times 86$ with compression factor $m=2$. The dimension of the
codebook and each codeword are \num{512} and \num{64}, respectively. The decoder has two $3 \times 3$ residual blocks, followed by two transposed convolutional layers with stride \num{2} and kernel size $4 \times 4$. We train the VQ-VAE model using the Adam optimizer \cite{kingma2014adam} with a learning rate of \num{3e-4} and a batch size of \num{64} for \num{70000} iterations.


\subsubsection{Autoregressive model}
The PixelSNAIL \cite{chen2018pixelsnail} model is trained on the $20 \times 86$ DTFR of sound using the Adam optimizer \cite{kingma2014adam} with a learning rate of \num{3e-4} and a batch size of \num{32} for \num{250000} iterations. We use a PyTorch implementation of PixelSNAIL\footnote{\url{https://github.com/rosinality/vq-vae-2-pytorch/blob/master/pixelsnail.py}}.

\subsubsection{Waveform synthesis module}
The generated mel-spectrograms are converted into waveforms using HiFi-GAN \cite{kong2020hifi}, which provides high-fidelity speech synthesis results and fast inference. We train a HiFi-GAN on the training data of UrbanSound8K dataset from scratch using the code provided in the official GitHub repository\footnote{\url{https://github.com/jik876/hifi-gan}}. 

\subsection{Baseline system}
SampleRNN has been adapted for sound generation in \cite{kong2019acoustic}. In this work, we use a two-tier conditional SampleRNN\footnote{\url{https://github.com/qiuqiangkong/sampleRNN_acoustic_scene_generation}} as the baseline system. The baseline system is trained on raw waveforms for \num{350000} iterations using the Adam optimizer \cite{kingma2014adam} with a learning rate of \num{1e-3} and a batch size of \num{64}.

\subsection{Evaluation methods}
Several subjective metrics \cite{okamoto2019overview} have been proposed for evaluating the performance of acoustic generative models. However, a subjective evaluation of sound is time-consuming and the results are sometimes difficult to reproduce. In this work, we adopt the quality and diversity of generated sound samples as two objective performance metrics.

\subsubsection{Generation quality}
Similar to the evaluation metric used in \cite{kong2019acoustic}, we train a VGG11 \cite{iqbal2018general} classifier on the training data and then use the trained VGG11 to classify the generated data. If the generated data is of high quality, the VGG11 will assign them to the corresponding sound classes with high accuracy. If the generated data is of low quality, such as random noise, the VGG11 will tend to predict them as random classes. Although this metric does not indicate the perceptual quality of the generated sound, it is still useful for partially assessing how good the generated sound is. The VGG11 classifier is trained on the computed spectrogram
(mentioned in Section 3.2) of training data using the Adam optimization algorithm \cite{kingma2014adam} with a batch size of \num{128} and a learning rate of \num{5e-4}. The VGG11 classifier achieves a \num{83}\% accuracy on testing data after training for \num{3100} iterations.

\subsubsection{Generation diversity}
The number of statistically-different bins (NDB) \cite{richardson2018gans} has been proposed to evaluate generative models. This evaluation metric first clusters the training data into different bins and then assigns each generated data to the nearest bin. NDB is reported as the number of bins where the number of training instances is statistically different from the number of generated instances by a two-sample Binomial test. In addition, the Jensen-Shannon divergence (JSD) between the distribution of the training data and generated data over the clustered bins is calculated as the evaluation metric if the number of samples is sufficiently large. A smaller NDB and JSD represent better performance.
We adopt the K-means algorithm to cluster sound data in the T-F domain (as reported in Section 3.2). We then calculate the NDB and JSD of the generated sound in the class-wise case and the all-classes case (merge the generated data of all classes together and compare with the training data), respectively. \num{20} bins are used for class-wise clustering and \num{200} bins are used for all-classes clustering. We use the official implementation of NDB and JSD\footnote{\url{https://github.com/eitanrich/gans-n-gmms/blob/master/utils/ndb.py}}.

\subsection{Evaluation results}
We use our proposed method and the baseline to generate \num{1024} sound clips per class. Evaluation results are discussed below.

\subsubsection{Generation quality}
Table \ref{tab:quality} shows a VGG11 classification accuracy of \num{52.89}\%, \num{52.81}\%, \num{82.97}\%, \num{80.74}\% on the data generated by our proposed approach (Proposed Approach), data
generated by baseline (SampleRNN), testing data (Test), and testing data after the reconstruction based on DTFR (Reconstructed Test), respectively. Our proposed approach achieves a comparable performance in generation quality compared with SampleRNN. Sound classes such as dog bark and gunshot perform better, while sound classes such as jackhammer and children playing perform worse. In addition, although the DTFR is four times smaller than the spectrogram, the classification accuracy on the testing data after reconstruction only decreases by \num{2.23} percentage points, which confirms the effectiveness of DTFR.

\subsubsection{Generation diversity}
The results of class-wise and all-classes evaluations of generation diversity are shown in Table \ref{tab:diversity_class} and Table \ref{tab:diversity_whole}, respectively. Our proposed approach outperforms the SampleRNN baseline significantly in NDB and JSD for all sound classes, which means the data generated by our approach has greater diversity and its distribution is closer to the real data. The JSD of the testing data is higher than the data generated by our proposed approach because the size of the testing data is small and the class distribution is different from the training data.

\begin{table}[]
\centering
\caption{Results of all-classes NDB and JSD}
\label{tab:diversity_whole}
\setlength{\tabcolsep}{5.5mm}{
\begin{tabular}{@{}ccc@{}}
\toprule
                  & NDB$_{\text{all-classes}}$ & JSD$_{\text{all-classes}}$ \\ \midrule
Proposed Approach & 25          & 0.0461          \\ \midrule
SampleRNN & 120          & 0.3267          \\ \midrule
Test         & 6          & 0.1359          \\ \bottomrule
\end{tabular}}
\end{table}

\subsection{Ablation study}
We investigate the impact of the multi-scale convolutional scheme (MSCS) in the VQ-VAE's encoder. Table \ref{tab:Ablations} shows the mean square error (MSE) and the VGG11 classification accuracy of the reconstructed test data based on DTFR with and without MSCS. Experimental results show that by applying the MSCS, the MSE decreases by \num{0.0047} and the VGG11 classification accuracy increases by \num{6.2} percentage points, which indicates that more acoustic information (i.e. local fine-grained structures) within sound is captured by MSCS.

\section{Conclusions}
\label{sec:conclusion}

We have presented a novel approach for conditional sound generation using neural discrete time-frequency representation learning. Our proposed approach can efficiently model long-range dependencies and retrain local fine-grained structures within sound clips. Experimental results show that our proposed method has better performance in diversity and has comparable performance in quality compared to SampleRNN. In future work, we will consider learning a representation via adversarial training \cite{makhzani2015adversarial} and perceptual loss \cite{liu2017perceptually}, and compare it with other GAN-based audio generative model \cite{donahue2018adversarial}.

\begin{table}[]
\centering
\caption{Results of ablation experiment}
\label{tab:Ablations}
\setlength{\tabcolsep}{8mm}{
\begin{tabular}{@{}ccc@{}}
\toprule
                  & MSE & Accuracy \\ \midrule
DTFR w/ \ \ MSCS & 0.0684          & 0.8074          \\ \midrule
DTFR w/o MSCS         & 0.0731          & 0.7454          \\ \bottomrule
\end{tabular}}
\end{table}

\bibliographystyle{IEEEtran}

\bibliography{refs}

\begin{thebibliography}{10}
\providecommand{\url}[1]{#1}
\csname url@samestyle\endcsname
\providecommand{\newblock}{\relax}
\providecommand{\bibinfo}[2]{#2}
\providecommand{\BIBentrySTDinterwordspacing}{\spaceskip=0pt\relax}
\providecommand{\BIBentryALTinterwordstretchfactor}{4}
\providecommand{\BIBentryALTinterwordspacing}{\spaceskip=\fontdimen2\font plus
\BIBentryALTinterwordstretchfactor\fontdimen3\font minus
  \fontdimen4\font\relax}
\providecommand{\BIBforeignlanguage}[2]{{%
\expandafter\ifx\csname l@#1\endcsname\relax
\typeout{** WARNING: IEEEtran.bst: No hyphenation pattern has been}%
\typeout{** loaded for the language `#1'. Using the pattern for}%
\typeout{** the default language instead.}%
\else
\language=\csname l@#1\endcsname
\fi
#2}}
\providecommand{\BIBdecl}{\relax}
\BIBdecl

\bibitem{virtanen2018computational}
T.~Virtanen, M.~D. Plumbley, and D.~Ellis, ``Introduction to sound scene and
  event analysis,'' in \emph{Computational Analysis of Sound Scenes and
  Events}, 1st~ed.\hskip 1em plus 0.5em minus 0.4em\relax Springer, 2018.

\bibitem{lloyd2011sound}
D.~B. Lloyd, N.~Raghuvanshi, and N.~K. Govindaraju, ``Sound synthesis for
  impact sounds in video games,'' in \emph{Symposium on Interactive 3D Graphics
  and Games}, 2011, pp. 55--62.

\bibitem{salamon2017scaper}
J.~Salamon, D.~MacConnell, M.~Cartwright, P.~Li, and J.~P. Bello, ``Scaper: A
  library for soundscape synthesis and augmentation,'' in \emph{2017 IEEE
  Workshop on Applications of Signal Processing to Audio and Acoustics
  (WASPAA)}, 2017, pp. 344--348.

\bibitem{ren2018deep}
Z.~Ren, K.~Qian, Z.~Zhang, V.~Pandit, A.~Baird, and B.~Schuller, ``Deep
  scalogram representations for acoustic scene classification,'' \emph{IEEE/CAA
  Journal of Automatica Sinica}, vol.~5, no.~3, pp. 662--669, Apr. 2018.

\bibitem{kong2020sound}
Q.~Kong, Y.~Xu, W.~Wang, and M.~D. Plumbley, ``Sound event detection of weakly
  labelled data with {CNN-transformer} and automatic threshold optimization,''
  \emph{IEEE/ACM Transactions on Audio, Speech, and Language Processing},
  vol.~28, pp. 2450--2460, Aug. 2020.

\bibitem{akkermans2011freesound}
V.~Akkermans, F.~Font~Corbera, J.~Funollet, B.~De~Jong, G.~Roma~Trepat,
  S.~Togias, and X.~Serra, ``Freesound 2: An improved platform for sharing
  audio clips,'' in \emph{Proceedings of the 12th International Society for
  Music Information Retrieval Conference (ISMIR)}, 2011.

\bibitem{shen2018natural}
J.~Shen, R.~Pang, R.~J. Weiss, M.~Schuster, N.~Jaitly, Z.~Yang, Z.~Chen,
  Y.~Zhang, Y.~Wang, R.~Skerrv-Ryan \emph{et~al.}, ``Natural {TTS} synthesis by
  conditioning {WaveNet} on mel spectrogram predictions,'' in \emph{IEEE
  International Conference on Acoustics, Speech and Signal Processing
  (ICASSP)}, 2018, pp. 4779--4783.

\bibitem{wang2017tacotron}
Y.~Wang, R.~Skerry-Ryan, D.~Stanton, Y.~Wu, R.~J. Weiss, N.~Jaitly, Z.~Yang,
  Y.~Xiao, Z.~Chen, S.~Bengio, Q.~V. Le, Y.~Agiomyrgiannakis, R.~Clark, and
  R.~A. Saurous, ``Tacotron: Towards end-to-end speech synthesis,'' in
  \emph{Proceedings of Interspeech}, 2017.

\bibitem{yang2017midinet}
L.-C. Yang, S.-Y. Chou, and Y.-H. Yang, ``{MidiNet}: A convolutional generative
  adversarial network for symbolic-domain music generation,'' in
  \emph{Proceedings of the 18th International Society for Music Information
  Retrieval (ISMIR)}, 2017.

\bibitem{dieleman2018challenge}
S.~Dieleman, A.~V.~D. Oord, and K.~Simonyan, ``The challenge of realistic music
  generation: modelling raw audio at scale,'' \emph{Advances in Neural
  Information Processing Systems (NIPS) 31}, pp. 8000--8010, 2018.

\bibitem{kong2019acoustic}
Q.~Kong, Y.~Xu, T.~Iqbal, Y.~Cao, W.~Wang, and M.~D. Plumbley, ``Acoustic scene
  generation with conditional {SampleRNN},'' in \emph{IEEE International
  Conference on Acoustics, Speech and Signal Processing (ICASSP)}, 2019, pp.
  925--929.

\bibitem{okamoto2021onoma}
Y.~Okamoto, K.~Imoto, S.~Takamichi, R.~Yamanishi, T.~Fukumori, and
  Y.~Yamashita, ``Onoma-to-wave: Environmental sound synthesis from
  onomatopoeic words,'' \emph{arXiv preprint arXiv:2102.05872}, 2021.

\bibitem{mehri2016samplernn}
S.~Mehri, K.~Kumar, I.~Gulrajani, R.~Kumar, S.~Jain, J.~Sotelo, A.~Courville,
  and Y.~Bengio, ``{SampleRNN}: An unconditional end-to-end neural audio
  generation model,'' \emph{5th International Conference on Learning
  Representations (ICLR)}, 2017.

\bibitem{vasquez2019melnet}
S.~Vasquez and M.~Lewis, ``{MelNet}: A generative model for audio in the
  frequency domain,'' \emph{arXiv preprint arXiv:1906.01083}, 2019.

\bibitem{kumar2019melgan}
K.~Kumar, R.~Kumar, T.~de~Boissiere, L.~Gestin, W.~Z. Teoh, J.~Sotelo,
  A.~de~Br{\'e}bisson, Y.~Bengio, and A.~Courville, ``{MelGAN}: Generative
  adversarial networks for conditional waveform synthesis,'' \emph{Advances in
  Neural Information Processing Systems (NIPS) 32}, 2019.

\bibitem{kong2020hifi}
J.~Kong, J.~Kim, and J.~Bae, ``{HiFi-GAN}: Generative adversarial networks for
  efficient and high fidelity speech synthesis,'' \emph{Advances in Neural
  Information Processing Systems (NIPS) 33}, 2020.

\bibitem{oord2017neural}
A.~V.~D. Oord, O.~Vinyals, and K.~Kavukcuoglu, ``Neural discrete representation
  learning,'' \emph{Advances in Neural Information Processing Systems (NIPS)
  30}, pp. 6306--6315, 2017.

\bibitem{chen2018pixelsnail}
X.~Chen, N.~Mishra, M.~Rohaninejad, and P.~Abbeel, ``{PixelSNAIL}: An improved
  autoregressive generative model,'' in \emph{International Conference on
  Machine Learning (ICML)}, 2018, pp. 864--872.

\bibitem{richardson2018gans}
E.~Richardson and Y.~Weiss, ``On {GANs} and {GMMs},'' \emph{Advances in Neural
  Information Processing Systems (NIPS) 31}, 2018.

\bibitem{salamon2014dataset}
J.~Salamon, C.~Jacoby, and J.~P. Bello, ``A dataset and taxonomy for urban
  sound research,'' in \emph{Proceedings of the 22nd ACM International
  Conference on Multimedia}, 2014, pp. 1041--1044.

\bibitem{xian2021multi}
Y.~Xian, Y.~Sun, W.~Wang, and S.~M. Naqvi, ``Multi-scale residual convolutional
  encoder decoder with bidirectional long short-term memory for single channel
  speech enhancement,'' in \emph{28th European Signal Processing Conference
  (EUSIPCO)}, 2021, pp. 431--435.

\bibitem{oord2016conditional}
A.~V.~D. Oord, N.~Kalchbrenner, O.~Vinyals, L.~Espeholt, A.~Graves, and
  K.~Kavukcuoglu, ``Conditional image generation with {PixelCNN} decoders,''
  \emph{Advances in Neural Information Processing Systems (NIPS) 29}, 2016.

\bibitem{vaswani2017attention}
A.~Vaswani, N.~Shazeer, N.~Parmar, J.~Uszkoreit, L.~Jones, A.~N. Gomez,
  L.~Kaiser, and I.~Polosukhin, ``Attention is all you need,'' \emph{Advances
  in Neural Information Processing Systems (NIPS) 30}, 2017.

\bibitem{kingma2014adam}
D.~P. Kingma and J.~Ba, ``Adam: A method for stochastic optimization,''
  \emph{3rd International Conference on Learning Representations (ICLR)}, 2015.

\bibitem{okamoto2019overview}
Y.~Okamoto, K.~Imoto, T.~Komatsu, S.~Takamichi, T.~Yagyu, R.~Yamanishi, and
  Y.~Yamashita, ``Overview of tasks and investigation of subjective evaluation
  methods in environmental sound synthesis and conversion,'' \emph{arXiv
  preprint arXiv:1908.10055}, 2019.

\bibitem{iqbal2018general}
T.~Iqbal, Q.~Kong, M.~D. Plumbley, and W.~Wang, ``General-purpose audio tagging
  from noisy labels using convolutional neural networks,'' in \emph{Proceedings
  of the Detection and Classification of Acoustic Scenes and Events}, 2018, pp.
  212--216.

\bibitem{makhzani2015adversarial}
A.~Makhzani, J.~Shlens, N.~Jaitly, I.~Goodfellow, and B.~Frey, ``Adversarial
  autoencoders,'' \emph{arXiv preprint arXiv:1511.05644}, 2015.

\bibitem{liu2017perceptually}
Q.~Liu, W.~Wang, P.~J. Jackson, and Y.~Tang, ``A perceptually-weighted deep
  neural network for monaural speech enhancement in various background noise
  conditions,'' in \emph{25th European Signal Processing Conference (EUSIPCO)},
  2017, pp. 1270--1274.

\bibitem{donahue2018adversarial}
C.~Donahue, J.~McAuley, and M.~Puckette, ``Adversarial audio synthesis,''
  \emph{7th International Conference on Learning Representations (ICLR)}, 2019.

\end{thebibliography}

\end{document}